\documentclass[12pt,a4paper]{article}
\usepackage[dvips]{graphicx}
\begin{document}
\begin{flushright}
LNF--00/032(P)\\
hep-ph/0012050\\
December 2000
\end{flushright}
\vspace*{0.8cm}

\begin{center}
{\Large\bf 
Chiral loop and L$\sigma$M predictions for $\phi\rightarrow\pi^0\eta\gamma$}\\
\vspace*{0.4cm}

R. Escribano\footnote{Talk given at the Hadron Structure'2000 Conference,
Stara Lesna, High Tatras Mountains, Slovakia, 2--7 October 2000.}\\
\vspace*{0.1cm}

{\footnotesize\it 
INFN-Laboratori Nazionali di Frascati, P.O. Box 13, I-00044 Frascati, Italy}\\
\end{center}
\vspace*{0.4cm}

\begin{abstract}
A prediction for the contributions of chiral loops and the 
L$\sigma$M model to the radiative $\phi\rightarrow\pi^0\eta\gamma$
decay mode is presented.
The L$\sigma$M is used as an appropriate framework for describing the
pole effects of the $a_0(980)$ scalar resonance.
As a result, a better agreement with present available data is achieved 
for the higher part of the $\pi^0\eta$ invariant mass spectrum.
For the branching ratio, a value of
$B(\phi\rightarrow\pi^0\eta\gamma)=(0.75$--$0.95)\times 10^{-4}$
is found.
\end{abstract}

\section{Introduction}
As exposed in detail by Dr.~Denig in his very nice and introductory 
talk on the KLOE status and first results 
(see A.~G.~Denig's talk in these proceedings), 
$\phi$ radiative decays into two pseudoscalar mesons and a photon are
very interesting processes to be measured.
Their accurate measurements could offer us the possibility of shedding 
some light on the presently unclear nature of scalars
(in particular, in those channels involving the $f_0(980)$ and $a_0(980)$
resonances).
These decays, and in general all the associated 
$V\rightarrow P^0 P^0\gamma$ radiative decays with $V=\rho, \omega, \phi$
and $P^0=\pi^0, K^0, \eta$, are not only a challenge for experimentalists 
but also for theorists. The reason for that is at least twofold.
First, the center of mass energy of these processes is around 1 GeV,
an energy region that is, on the one side, too low for perturbative QCD to
be applicable and, on the other, too high for Chiral Perturbation Theory
(ChPT) predictions to be reliable.
Accordingly, these processes provide us with an excellent laboratory for 
testing our knowledge of hadron physics in the 1 GeV energy domain.
Second, due to the quantum numbers of the initial vector and of the
final photon, both with $J^{PC}=1^{--}$, the remaining two pseudoscalar
meson system is in a $J^{PC}=0^{++}$ 
configuration\footnote{Rescattering effects from $2^{++}$ states are
very suppressed because the nearest tensorial resonances, $f_2(1270)$ 
and $a_2(1320)$, are well above the lightest vector 
masses \protect\cite{PDG}.}, 
whose quantum numbers correspond precisely to those of a scalar state.
So then, these vector radiative decays may help us to establish the 
nature of these scalar resonances, 
nowadays very controversial\footnote{Indeed, 
several proposals have been suggested along the years concerning the 
constitution of these scalars as 
multiquark states \cite{jaffe}, 
$K \bar K$ molecules \cite{weinstein} or 
ordinary $q \bar q$ mesons \cite{tornqvist:1}.} \cite{close},
and their poorly known properties (masses and decay widths) \cite{PDG},
thus adding further interest but also complexity to this 1 GeV energy
region.

In this presentation, we propose the linear sigma model (L$\sigma$M)
as an appropriate framework for incorporating the lightest scalar 
resonances and their pole effects into a ChPT inspired context.
This will allow us to benefit from the common origin of ChPT and the 
L$\sigma$M to improve the chiral loop predictions for the 
$V\rightarrow P^0 P^0\gamma$ decays exploiting the complementarity 
of both approaches.
As well known, ChPT is the established theory of the pseudoscalar
interactions at low energy. 
However, it is not reliable at energies of a typical vector meson mass 
and scalar resonance poles are not explicitly included. 
As a result, ChPT inspired loop models can give rough estimates for 
$B(V\rightarrow P^0 P^0\gamma)$ but will hardly be able to reproduce the 
observed enhancements in the invariant mass spectra. 
The L$\sigma$M is a much simpler model dealing similarly with 
pseudoscalar interactions but taking into account scalar resonances 
in a systematic and definite way. 
Consequently, the L$\sigma$M should be able to reproduce the resonance 
peaks in the spectra associated to the $0^{++}$ states and, 
although it does not provide a complete framework for the pseudoscalar
meson physics, this model might be of relevance in describing the scalar 
resonance effects when linked to a well established ChPT context.

The Novosibirsk CMD-2 and SND Collaborations have reported very recently
the branching ratio and the $\pi^0\eta$ invariant mass distribution 
for the $\phi\rightarrow\pi^0\eta\gamma$ decay. 
For the branching ratio, the CMD-2 Collaboration reports 
$B(\phi\rightarrow\pi^0\eta\gamma)=
(0.90\pm 0.24\pm 0.10)\times 10^{-4}$ \cite{CMD-2}, 
while the SND result is, consistently,  
$B(\phi\rightarrow\pi^0\eta\gamma)=
(0.88\pm 0.14\pm 0.09)\times 10^{-4}$ \cite{SND:1}.
The observed invariant mass distribution shows a significant enhancement 
at large $\pi^0\eta$ invariant mass that, according to 
Refs.~\cite{CMD-2,SND:1}, could be interpreted as a manifestation of a 
sizeable contribution of the $a_0(980)\gamma$ intermediate state.
The last issue of the Review of Particle Physics,
including the previous CMD-2 value together with an older SND
measurement \cite{SND:2} announces 
$B(\phi\rightarrow\pi^0\eta\gamma)=
(0.86\pm 0.18)\times 10^{-4}$ \cite{PDG},
while the preliminary result presented in this conference by the 
KLOE Collaboration is 
$B(\phi\rightarrow\pi^0\eta\gamma)=
(0.77\pm 0.15\pm 0.10)\times 10^{-4}$ (see A.G. Denig's contribution).
This and other radiative $\phi$ decays are intensively investigated 
at the Frascati $\phi$-factory DA$\Phi$NE \cite{daphne95:juliet}.

On the theoretical side, the $V \rightarrow P^0 P^0\gamma$ decays have been 
extensively studied \cite{close,achasov,bramon:1,lucio:1,oset}.
Specifically, it has been shown that the intermediate vector meson 
contributions to $\phi\rightarrow\pi^0\eta\gamma$ lead to a small
$B(\phi\rightarrow\pi^0\eta\gamma)_{\rm VMD}=5.4 \times 10^{-6}$ \cite{BGP:1}, 
whereas a chiral loop model closely linked to standard ChPT predicts 
$B(\phi\rightarrow\pi^0\eta\gamma)_\chi=3.0 \times 10^{-5}$ \cite{BGP:2}. 
Needless to say, the scalar resonance effects and, in particular, 
the resonance pole associated to the $a_0$(980) were not contemplated in 
these two schemes. 
The recent experimental data from Novosibirsk and Frascati 
---for both the branching ratio and 
the $\pi^0\eta$ invariant mass spectrum--- 
seem therefore to disfavour these predictions based on vector meson exchange 
and/or a simple extrapolation of ChPT ideas.
If we rely on the resonance picture, 
it is clear that the $a_0(980)$ scalar meson ---lying just below the $\phi$ 
mass and having the appropriate quantum numbers--- 
should play an important r\^ole in the $\phi\rightarrow\pi^0\eta\gamma$ decay. 
Several theoretical attempts to describe the effects of scalars in $\phi$ 
radiative decays have appeared so far: 
the ``no structure" model \cite{bramon:2},
the $K^+K^-$ model \cite{close,lucio:1}, 
where the $\phi\rightarrow a_0\gamma$ amplitude is generated through a loop 
of charged kaons, and the chiral unitary approach $(U\chi PT)$ \cite{oset}, 
where the decay $\phi\rightarrow\pi^0\eta\gamma$ occurs through a loop of 
charged kaons that subsequently annihilate into $\pi^0\eta\gamma$.
In the two former cases the scalar resonances are included {\it ad hoc}  
while in the latter they are generated dynamically by unitarizing the 
one-loop amplitudes.

As an example of our work, we discuss the 
$\phi\rightarrow\pi^0\eta\gamma$ decay mode.
This process is the easiest to calculate since it only involves 
the $a_0(980)$ scalar resonance whose properties are quite well known.
This circumstance reduces to a minimum the uncertainties of the calculation 
and will render our theoretical prediction more definite and solid.

This presentation is based on a more extensive analysis published in 
Ref.~\protect\cite{paper}.

\section{Chiral loop prediction}
\label{sectChPT}
The vector meson initiated $V\rightarrow P^0 P^0\gamma$ decays cannot be 
treated in strict ChPT. 
This theory has to be extended to incorporate on-shell vector meson fields. 
At lowest order, this may be easily achieved by means of the 
${\cal O}(p^2)$ ChPT Lagrangian
\begin{equation}
\label{ChPTlag}
{\cal L}_2=\frac{f^2}{4}
\langle D_\mu U^\dagger D^\mu U+M(U+U^\dagger)\rangle\ ,
\end{equation}
where $f=f_\pi=92.4$ MeV at this order, 
$U=\exp(i\sqrt{2}P/f)$ with $P$ being the usual pseudoscalar nonet matrix, 
and $M=\mbox{diag}(m_\pi^2, m_\pi^2, 2 m_K^2-m_\pi^2)$ in the isospin limit.
The covariant derivative, now enlarged to include vector mesons, is defined as 
$D_\mu U=\partial_\mu U -i e A_\mu [Q,U] - i g [V_\mu,U]$, with 
$Q=\mbox{diag}(2/3, -1/3, -1/3)$ being the quark charge matrix and 
$V_\mu$ the additional matrix containing the nonet of ideally mixed vector 
meson fields. 
The diagonal elements of $V$ are 
$(\rho^0+\omega)/\sqrt{2}, (-\rho^0+\omega)/\sqrt{2}$ and $\phi$.

There is no tree-level contribution from this Lagrangian to the  
$\phi\rightarrow\pi^0\eta\gamma$ amplitude and at the one-loop level 
one needs to compute the set of diagrams shown in Ref.~\cite{BGP:2}.
A straightforward calculation leads to the following finite amplitude for 
$\phi(q^\ast,\epsilon^\ast)\rightarrow 
\pi^0(p)\eta(p^\prime)\gamma(q,\epsilon)$:
\begin{equation}
\label{AphiChPT}
{\cal A}(\phi\rightarrow \pi^0\eta\gamma)_\chi = \frac{eg}{2\pi^2 m^2_{K^+}}
(\epsilon^\ast\epsilon\ q^\ast q-\epsilon^\ast q\ \epsilon q^\ast)
L(m^2_{\pi^0\eta}){\cal A}(K^+K^-\rightarrow\pi^0\eta)_\chi\ ,
\end{equation}
where $L(m^2_{\pi^0\eta})$ is the loop integral function and
$m^2_{\pi^0\eta}$ is the invariant mass of the final pseudoscalar system
(see Ref.~\cite{paper} for further details).
The coupling constant $g$ is defined through the strong amplitude
${\cal A}(\phi\rightarrow K^+K^-)=g\epsilon^\ast (p_+-p_-)$ and
is the part beyond standard ChPT which is fixed from phenomenology.
The four-pseudoscalar amplitude is instead a standard ChPT 
amplitude\footnote{${\cal A}(K^+K^-\rightarrow\pi^0\eta_8)_\chi=
\frac{\sqrt{3}}{4 f_\pi^2}\left(m^2_{\pi^0\eta}-\frac{4}{3}m^2_K\right)$ 
if only the $\eta_8$ contribution is taken into account as 
in Ref.~\protect\cite{BGP:2}.}
which is found to be
\begin{equation}
\label{A4PChPTphys}
{\cal A}(K^+K^-\rightarrow\pi^0\eta)_\chi=\frac{1}{\sqrt{6}f_\pi^2}
\left(m^2_{\pi^0\eta}-\frac{10}{9}m^2_K+\frac{1}{9}m^2_\pi\right)\ .
\end{equation}
In Eqs.~(\ref{AphiChPT}, \ref{A4PChPTphys}), a value of
$\theta_P=\arcsin(-1/3)\simeq -19.5^\circ$ is taken for the
$\eta$-$\eta^\prime$ mixing angle.
This value, based on classical arguments of nonet symmetry,
is in fairly agreement with recent phenomenological 
estimates \cite{thetaP}.

\begin{figure}
\centerline{\includegraphics{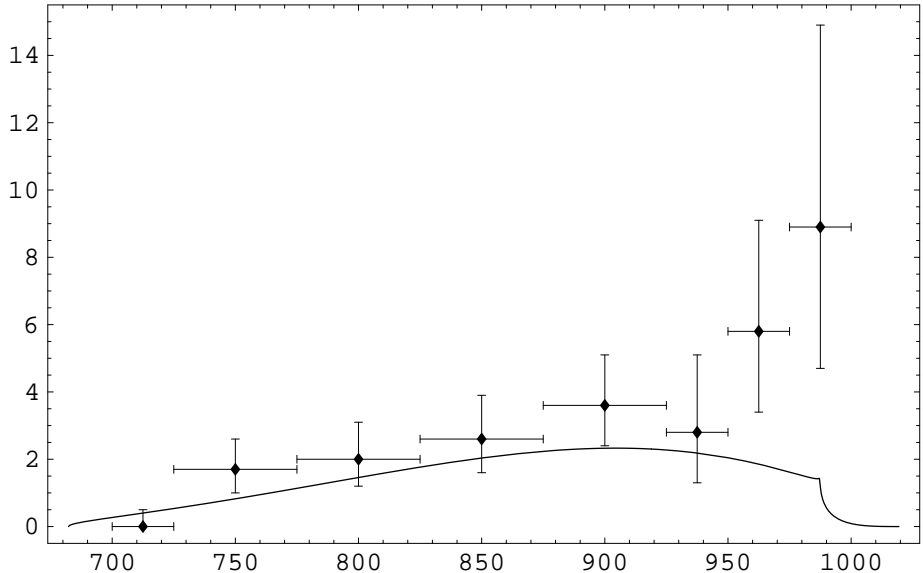}}
\caption{$dB(\phi\rightarrow\pi^0\eta\gamma)/dm_{\pi^0\eta}
\times 10^7\ \mbox{MeV}^{-1}$ as a function of the $m_{\pi^0\eta}$ 
invariant mass in a chiral loop model. 
Experimental data are taken from Ref.~\protect\cite{SND:1}.}
\label{figChPT}
\end{figure}
 
The invariant mass distribution for the $\phi\rightarrow\pi^0\eta\gamma$ 
decay is then given by the spectrum 
(see Fig.~\ref{figChPT})
\begin{equation}
\label{dGChPT}
\begin{array}{rl}
\frac{d\Gamma(\phi\rightarrow\pi^0\eta\gamma)_\chi}{dm_{\pi^0\eta}}&=\ 
\frac{\alpha}{192\pi^5}\frac{g^2}{4\pi}\frac{m^4_\phi}{m^4_{K^+}}
\frac{m_{\pi^0\eta}}{m_\phi}\left(1-\frac{m^2_{\pi^0\eta}}{m^2_\phi}\right)^3
\sqrt{1-2\frac{m^2_{\pi^0}+m^2_\eta}{m^2_{\pi^0\eta}}+
\left(\frac{m^2_\eta-m^2_{\pi^0}}{m^2_{\pi^0\eta}}\right)^2}\\[2ex]
&\times\ |L(m^2_{\pi^0\eta})|^2
|{\cal A}(K^+K^-\rightarrow\pi^0\eta)_\chi|^2\ .
\end{array}
\end{equation}
Integrating Eq.~(\ref{dGChPT}) over the whole physical region 
one obtains for the branching ratio
\begin{equation}
\label{BRChPT}
B(\phi\rightarrow\pi^0\eta\gamma)_\chi=0.47\times 10^{-4}\ .
\end{equation}
As expected, Fig.~\ref{figChPT} shows that our chiral loop approach 
gives a reasonable prediction for the lower part of the spectrum but 
fails to reproduce the observed enhancement in its higher part, 
where $a_0(980)$-resonance effects should manifest.
As a consequence, the branching ratio is below the experimental value 
by about a factor of 2.

\section{L$\sigma$M improved prediction}
\label{sectLsigmaM}
To analyze the scalar resonance effects in the $V\rightarrow P^0 P^0\gamma$ 
decay amplitudes, we use the linear sigma model 
(L$\sigma$M) \cite{oldsigmamodel} 
as an adequate framework for describing such effects.
It is a well-defined $U(3)\times U(3)$ chiral model which incorporates 
{\it ab initio} both the nonet of pseudoscalar mesons together with its 
chiral partner, the scalar mesons nonet.
Recently, this model has shown to be rather phenomenologically 
successful in studying the implications of chiral symmetry for the 
controversial scalar sector of QCD \cite{napsuciale,tornqvist:2,lucio:2}.

In this context, the $V\rightarrow P^0 P^0\gamma$ decays proceed through a 
loop of charged pseudoscalar mesons emitted by the initial vector that,
due to the additional emission of a photon, can rescatter into pairs of
neutral pseudoscalars with the quantum numbers of a scalar state.
The scalar resonances are then expected to play an essential r\^ole in this 
rescattering process and the L$\sigma$M will be shown particularly 
appropriate to fix the corresponding amplitudes.

For the case of $\phi\rightarrow\pi^0\eta\gamma$, the dominant contribution 
arises exclusively from a loop of charged kaons\footnote{A loop of charged
pions is highly suppressed because it involves the isospin violating and 
OZI--rule forbidden $\phi\pi\pi$ coupling.}
that subsequently rescatters into the final $J^{PC} = 0^{++}$ 
$\pi^0\eta$ state.
In the L$\sigma$M, the $K^+K^-\rightarrow\pi^0\eta$ rescattering
amplitude is driven by a contact term, a term with an $a_0$ exchanged in the 
$s$-channel, and two terms with a $\kappa$ 
({\it i.e.~}the strange $I=1/2$ scalar resonance) exchanged in the $t$- and 
$u$-channels. 
However, the latter $\kappa$-exchange contributions are absent for an 
$\eta$-$\eta^\prime$ mixing angle $\theta_P=\arcsin(-1/3)\simeq -19.5^\circ$ 
since the $g_{\kappa K\eta}$ coupling constant appearing in one of the 
$\kappa$ vertices vanishes\footnote{See Ref.~\protect\cite{paper} for a 
detailed discussion on the effect of neglecting these 
$\kappa$ contributions.}. 

A straightforward calculation of the $\phi\rightarrow\pi^0\eta\gamma$ decay 
amplitude leads to an expression identical to that in Eq.~(\ref{AphiChPT}) 
but with the four-pseudoscalar amplitude now computed in a L$\sigma$M context. 
In this case, the amplitude is just
\begin{equation}
\label{A4PLsigmaMphys}
{\cal A}(K^+K^-\rightarrow\pi^0\eta)_{\mbox{\scriptsize L$\sigma$M}}= 
\frac{1}{\sqrt{6}f_Kf_\pi}(m^2_{\pi^0\eta}-m^2_K)\times
\frac{m^2_\eta-m^2_{a_0}}{D_{a_0}(m^2_{\pi^0\eta})}\ ,
\end{equation}
where $D_{a_0}(m^2_{\pi^0\eta})$ is the $a_0$ propagator.

It is worth mentioning a few remarks on the four-pseudoscalar L$\sigma$M
amplitude in Eq.~(\ref{A4PLsigmaMphys}) and its ChPT counterpart in 
Eq.~(\ref{A4PChPTphys}):
{\it i)} for $m_{a_0}\rightarrow\infty$ and ignoring $SU(3)$-breaking in 
the pseudoscalar masses and decay constants, 
the L$\sigma$M amplitude reduces to the ChPT one.
This means that the large scalar mass limit of the L$\sigma$M mimics
perfectly (in the $SU(3)$ limit) the contributions from the derivative and
massive terms of the ChPT Lagrangian (\ref{ChPTlag}).
This, we believe, is the main virtue of our approach and makes the use of 
the L$\sigma$M reliable;
{\it ii)} the reason for the L$\sigma$M and ChPT yielding slightly
different amplitudes in the $m_{a_0}\rightarrow\infty$ limit is because
of the way $SU(3)$-symmetry is broken in the two approaches.
While in the L$\sigma$M a non $SU(3)$ symmetric choice of the vacuum 
expectation values makes simultaneously $m_\pi^2\neq m_K^2$ and 
$f_\pi\neq f_K$ \cite{napsuciale,tornqvist:2}, in ChPT $m_\pi^2\neq m_K^2$ 
is already present at tree level whereas $f_\pi\neq f_K$ is only achieved 
at next-to-leading order;
{\it iii)} the vicinity of the $a_0(980)$ mass to the $\pi^0\eta$
production threshold makes unavoidable the presence of the $a_0$
propagator to take into account the pole effects.
Its inclusion should guarantee the proper behavior of the higher part
of the $\pi^0\eta$ invariant mass spectrum;
{\it iv)} concerning the $a_0$ propagator, the opening of the $K\bar{K}$
channel close to the $a_0(980)$ mass generates some uncertainties about
which precise form for that propagator should be used.
A first possibility consists in using a Breit-Wigner propagator with an 
energy dependent width to incorporate the known kinematic corrections:
$D_{a_0}(s)=s-m^2_{a_0}+i\sqrt{s}\,\Gamma_{a_0}(s)$,
where $\Gamma_{a_0}(s)=\sum_{ab}\Gamma_{a_0}^{ab}(s)$ and
\begin{equation}
\label{GammaBW}
\Gamma_{a_0}^{ab}(s)=
\frac{g^2_{a_0 ab}}{16\pi\sqrt{s}} 
\sqrt{\left[1-\frac{(m_a+m_b)^2}{s}\right]
      \left[1-\frac{(m_a-m_b)^2}{s}\right]}\, 
\theta (\sqrt{s}-(m_a+m_b))\ ,
\end{equation}
for $ab=\pi^0\eta, K^+ K^-, K^0 \bar{K}^0$
(see Ref.~\cite{paper} for the coupling constants $g_{a_0 ab}$).
Another interesting and widely accepted option was proposed by Flatt\'e
time ago specifically to the two-channel $a_0$ resonance \cite{flatte}:
$D_{a_0}(s)=s-m^2_{a_0}+im_{a_0}\,\Gamma_{a_0}(s)$,
where $\Gamma_{a_0}(s)=
\Gamma_{a_0}^{\pi^0\eta}(s)+\Gamma_{a_0}^{K\bar{K}}(s)$ and
\begin{equation}
\label{Gammaflatte}
\Gamma_{a_0}^{K\bar{K}}(s)=
\frac{g^2_{a_0 K\bar{K}}}{16\pi\sqrt{s}}\times
\left\{
\begin{array}{ll}
\sqrt{1-\frac{4m^2_K}{s}} &\ \mbox{for $\sqrt{s}\geq 2m_K$}\\[2ex]
i\sqrt{\frac{4m^2_K}{s}-1} &\ \mbox{for $\sqrt{s}< 2m_K$}
\end{array}
\right.
\end{equation}
for $K\bar{K}=K^+ K^-, K^0 \bar{K}^0$.
The relative narrowness of the observed ${\pi\eta}$ peak around 980 MeV 
is then explained by the action of unitarity and analyticity at the 
$K\bar{K}$ threshold.

Due to these distinct possibilities to deal with the $a_0$ propagator, 
as well as other differences introduced when implementing and fitting 
the basic L$\sigma$M Lagrangian by several authors, a set of predictions
can be obtained for the four-pseudoscalar amplitude (\ref{A4PLsigmaMphys}). 
In turn, these various amplitudes have to substitute the four-pseudoscalar 
ChPT amplitude in Eq.~(\ref{dGChPT}) to finally obtain the corresponding 
invariant mass distributions of the $\phi\rightarrow\pi^0\eta\gamma$ 
decay mode.  

\begin{figure}
\centerline{\includegraphics{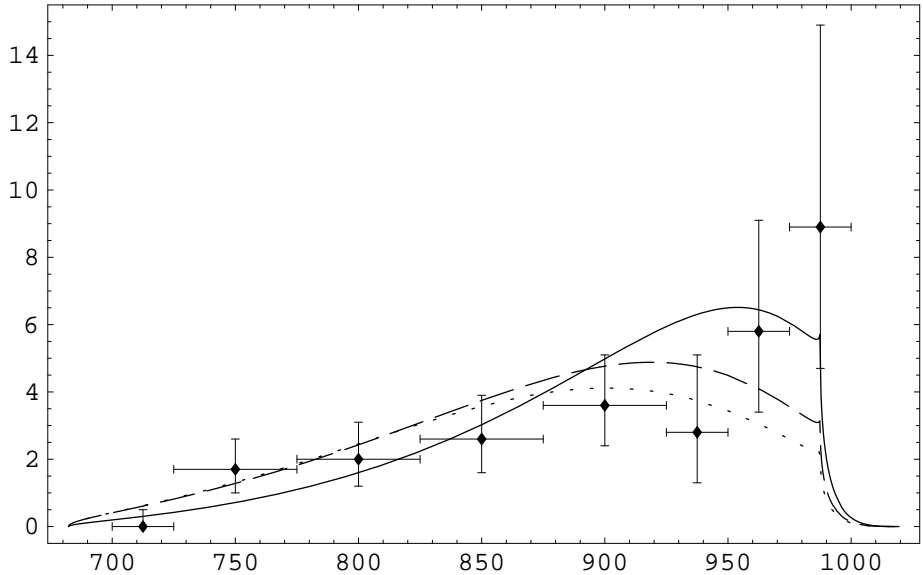}}
\caption{$dB(\phi\rightarrow\pi^0\eta\gamma)/dm_{\pi^0\eta}\times 
10^7\ \mbox{MeV}^{-1}$ as a function of the $m_{\pi^0\eta}$ invariant mass 
in the L$\sigma$M.
The dotted, dashed and solid lines correspond to the versions of the 
L$\sigma$M proposed by Refs.~\protect\cite{napsuciale,tornqvist:2,shabalin} 
respectively. 
Experimental data are taken from Ref.~\protect\cite{SND:1}.}
\label{figLsigmaM}
\end{figure}

We start our discussion along the lines of Ref.~\cite{napsuciale} taking 
for the $a_0$ propagator the simple Breit-Wigner prescription. 
The use of this propagator for the L$\sigma$M 
amplitude Eq.~(\ref{A4PLsigmaMphys}) and its insertion in Eq.~(\ref{dGChPT})
predicts the $m_{\pi^0\eta}$ invariant mass spectrum shown by the 
dotted line in Fig.~\ref{figLsigmaM}. 
Integrating over the whole physical region leads to the branching ratio
\begin{equation}
\label{BRmauro}
B(\phi\rightarrow\pi^0\eta\gamma)_
{\mbox{\scriptsize L$\sigma$M\cite{napsuciale}}}=
0.80\times 10^{-4}\ ,
\end{equation}
in agreement with the experimental branching ratio.
However, since the simple expression used for the $a_0$ propagator implies 
a large $a_0$-width 
($\Gamma_{a_0\rightarrow\pi\eta}\simeq 460$ MeV \cite{napsuciale}), 
the desired enhancement in the invariant mass spectrum appears in its
central part rather than around the $a_0$ peak.

This unpleasant feature is partially corrected when turning to the proposal 
by T\"ornqvist \cite{tornqvist:2}, where a Gaussian form factor related to 
the finite size of physical mesons and depending on the final CM-momentum 
is introduced to describe the decays of scalar resonances in this approach.
As a result, the decay width of $a_0$(980) into $\pi^0\eta$ is reduced 
($\Gamma_{a_0\rightarrow\pi\eta}\simeq 273$ MeV \cite{tornqvist:2}) 
without affecting that of $a_0$(980) into $K\bar{K}$. 
This fact produces an enhancement in the spectrum for the higher values of 
the $m_{\pi^0\eta}$ invariant mass, as shown by the dashed line  
in Fig.~\ref{figLsigmaM}. 
The integrated branching ratio is then predicted to be 
\begin{equation}
\label{BRniels}
B(\phi\rightarrow\pi^0\eta\gamma)_
{\mbox{\scriptsize L$\sigma$M\cite{tornqvist:2}}}=
0.90\times 10^{-4}\ .
\end{equation}

None of these drawbacks are encountered in the treatment proposed by 
Shabalin \cite{shabalin}, where the Flatt\'e corrections 
(indeed, a more precise form of them) are introduced in the $a_0$ propagator.
The $a_0$ width is then drastically reduced to a more acceptable visible 
width of $\Gamma_{a_0}\simeq 65$ MeV.
Within this approach, our prediction for the $m_{\pi^0\eta}$ invariant mass 
is shown by the solid line in Fig.~\ref{figLsigmaM} and the integrated 
branching ratio is 
\begin{equation}
\label{BRevgenii}
B(\phi\rightarrow\pi^0\eta\gamma)_
{\mbox{\scriptsize L$\sigma$M\cite{shabalin}}}=0.93\times 10^{-4}\ .
\end{equation}
Both the spectrum and the branching ratio are in nice agreement with the 
experimental data \cite{SND:1}.
The fact that Shabalin's treatment incorporates the Flatt\'e corrections 
to the $a_0$ resonant shape has played a major r\^ole in this achievement.

\section{Conclusions}
In this presentation, I have discussed a new amplitude for the
$\phi\rightarrow\pi^0\eta\gamma$ process that includes the effects of the
$a_0(980)$ scalar resonance.
The L$\sigma$M has shown to be a very suitable framework for including
the $a_0$ pole effects in a systematic and definite way.
An explanation of the higher part of the $\pi^0\eta$ invariant mass spectrum 
is then achieved. 
In the low invariant mass region, the result coincides with that coming from 
a chiral loop model calculation, thus making our whole approach reliable.
As a result of our analysis, it is safely concluded that the
$\phi\rightarrow\pi^0\eta\gamma$ branching ratio is in the range
$B(\phi\rightarrow\pi^0\eta\gamma)=(0.75$--$0.95)\times 10^{-4}$,
a prediction that is compatible with the present experimental data.
Nevertheless, the uncertainties affecting these predictions suggest that
more refined analyses are needed, particularly when the higher accuracy 
data from DA$\Phi$NE will be available.

\section*{Acknowledgements}
I would like to express my gratitude to the Hadron Structure'2000
Organizing Committee for the opportunity of presenting this contribution,
and for the pleasant and interesting conference we have enjoyed.

Work supported by the EEC, TMR-CT98-0169, EURODAPHNE network.

\end{document}